# Coupled mode theory of stock price formation

**Jack Sarkissian**

Managing Member, Algostox Trading LLC

641 Lexington Avenue, 15th floor, New York, NY 10022

email: jack@algostox.com

Abstract

We develop a theory of bid and ask price dynamics where the two prices form due to interaction of buy and sell orders. In this model the two prices are represented by eigenvalues of a 2x2 price operator corresponding to "bid" and "ask" eigenstates. Matrix elements of price operator fluctuate in time which results in phase jitter for eigenstates. We show that the theory reflects very important characteristics of bid and ask dynamics and order density in the order book. Calibration examples are provided for stocks at various time scales. Lastly, this model allows to quantify and measure risk associated with spread and its fluctuations.

## 1. The market maker's problem

Market making is an activity in which a company quotes buy and sell prices of financial instruments for other market participants, while providing a commitment to buy and sell at the quoted prices. The company's profit comes from the bid-ask spread between the buy and sell prices.

Basic principles of market making operations are easy to understand. When market maker's quote is crossed by a counterparty's order, it is executed and the market maker opens a position (called *induced inventory*). It is the market maker's purpose to close this position at the opposite price as soon as possible before the price moves in an unfavorable direction. Induced inventory is a subject to market fluctuations (*residual risk*) and is an unwanted component. When priced properly, the difference between bid and ask quotes (*spread*) provided by a market maker is supposed to cover that risk. The spread must also provide certain premium as a reward for bearing the residual risk.

In order to address the pricing problem one has to understand the behavior of order book, behavior of bid and ask prices, and the components making up the spread. A number of models were proposed to address the problem. Some of them are based on modeling order flow, others focus on price fluctuations [1-7].

These works treat price formation as a classical process with certain statistical properties. We propose a different approach, where price formation is treated as a quantum process, in the sense that price is described by a spectrum of values simultaneously and forms only as a result of measurement. Unless an order is executed, the asset can exist in a number of states with different prices represented by eigenvalues of an operator (the price operator). Price spectrum constantly changes due to the stochastic nature of the price operator.

Similarity of price formation with the eigenvalue problem is brought up by the behavior of order book data. First, formation of bid and ask prices can be thought of as a result of dynamic interaction between order levels in the order book. Such processes can be described by coupled-mode equations, in which eigenvalues represent prices and eigenfunctions represent order density at each level. Secondly, bid and ask price levels do not cross (in a single order book). They share this property with the result of Wigner-von-Neumann theorem, according to which eigenvalues also don't cross. And lastly, if matrix elements are allowed to fluctuate, so would the eigenvalues resulting in statistical distribution of order book data. This last fact will also explain absence of noticeable interference effects between the states. Fluctuation of matrix elements results in phase jitter and washes out any patterns.

The essential difference between classical and quantum descriptions has been extensively addressed in the literature [8, 9]. We will just mention the major relevant points. Classical approach assumes that price always exists and its value depends on assumptions about market participant behavior. In quantum approach all one can know is the price spectrum and the probabilities associated with each price. These probabilities do not necessarily result from market participants' behavior, but are intrinsic properties of price formation act. They cannot be fully described by statistical, kinetic, game-theoretical, or any classical-based models. Still, let us not forget that in many cases quantum systems can behave like classical ones.

Without loss of generality we will discuss equities here, although the theory can be applied to many other asset classes.

## 2. Bid and ask prices as eigenvalues of price operator

We will begin with declaring that the probabilities will be described by probability amplitudes, whose modulus squared represents the probability density itself:

$$p = |\psi|^2 .$$

Stock prices are governed by the *price operator* $\hat{S}$, whose eigenvalues represent the spectrum of prices, that the stock can attain:

$$\hat{S}\psi_n = s_n\psi_n . \tag{1}$$

Here

(a) price operator $\hat{S}$ naturally must have Hermitian properties since price is a real number

(b) eigenstates characterize the probability of finding the stock in state with price $s_n$ such that $p_n = |\psi_n|^2 .$

(c) eigenvalues represent stock price in the corresponding state

Price fluctuations, observed in financial markets are now included through fluctuations of the operator $\hat{S}$ :

$$\hat{S}(t + \delta t) = \hat{S}(t) + \delta\hat{S}(t)$$

Let us consider a model, in which the stock has only two states: one with price equal to bid price, and the other with price equal to ask price.

$$\psi = \begin{pmatrix} \psi_{ask} \\ \psi_{bid} \end{pmatrix},$$

Writing Eq. (1) in matrix form we have:

$$\begin{pmatrix} s_{11} & s_{12} \\ s_{12}^* & s_{22} \end{pmatrix} \begin{pmatrix} \psi_{ask} \\ \psi_{bid} \end{pmatrix} = s_{ask/bid} \begin{pmatrix} \psi_{ask} \\ \psi_{bid} \end{pmatrix} \tag{2}$$

Eigenvalues $s_{ask}$ and $s_{bid}$ are then expressed through matrix elements as

$$s_{ask} = \frac{s_{11} + s_{22}}{2} + \sqrt{\left(\frac{s_{11} - s_{22}}{2}\right)^2 + |s_{12}|^2} \qquad (3a)$$

$$s_{bid} = \frac{s_{11} + s_{22}}{2} - \sqrt{\left(\frac{s_{11} - s_{22}}{2}\right)^2 + |s_{12}|^2} \qquad (3b)$$

Here the first term represents the mid price and the second term is the semispread.

$$s_{mid} = \frac{s_{bid} + s_{ask}}{2} = \frac{s_{11} + s_{22}}{2} \qquad \text{(mid-price)} \qquad (4a)$$

$$\Delta = \sqrt{(s_{11} - s_{22})^2 + 4|s_{12}|^2} \qquad \text{(spread)} \qquad (4b)$$

With these notations Eqs. (3a, 3b) rewrite as

$$s_{ask} = s_{mid} + \frac{\Delta}{2} \qquad \text{and} \qquad s_{bid} = s_{mid} - \frac{\Delta}{2} \qquad (5)$$

## 3. Spread and its statistical properties

In order to apply this formalism to model bid and ask prices let us introduce fluctuations to matrix elements of the price operator:

$$s_{11}(t + dt) = s_{mid}(t) + \sigma dz + \frac{\xi}{2} \qquad (6a)$$

$$s_{22}(t + dt) = s_{mid}(t) + \sigma dz - \frac{\xi}{2} \qquad (6b)$$

$$s_{12}(t + dt) = \frac{\kappa}{2}, \qquad (6c)$$

where $\xi$ and $\kappa$ are normally distributed around their means, so at any point in time we can write:

$$\xi = \xi_0 + \xi_1 \, du \qquad \text{and} \qquad \kappa = \kappa_0 + \kappa_1 \, dv$$

Variables $z$, $u$, and $v$ are (generally) uncorrelated random variables with standard normal distribution. In such setup the mid price and the spread are given by equations:

$$s_{mid}(t + dt) = s_{mid}(t) + \sigma dz \qquad (7a)$$

$$\Delta = \sqrt{(\xi_0 + \xi_1 du)^2 + (\kappa_0 + \kappa_1 dv)^2} \qquad (7b)$$

As we can see, the mid price simply follows a Gaussian process with volatility $\sigma$. Statistics of spread is more complex and deserves a closer look. One can see that spread is made up of a number of components:

(a) the intrinsic component $\xi_0$ which exists even in the absence of market fluctuations and level interactions

(b) the interaction component $\kappa_0$ which is maintained by steady interaction between the levels

(c) risk components, associated with the fluctuations of the first two components.

Let us consider the case where $\xi_0 = 0$ and $\kappa_0 = 0$, which allows an analytical solution. It is a reasonable approximation in some cases, though in many cases the $\kappa_0$ is still substantial. In this case spread reduces to:

$$\Delta = \sqrt{(\xi_1 du)^2 + (\kappa_1 dv)^2} \qquad (8)$$

Probability distribution of this $\Delta$ is described by equation:

$$P(\Delta) = \frac{1}{\xi_1 \kappa_1} \Delta\, e^{-a\Delta^2} I_0(b\Delta^2) \qquad (9)$$

where

$$a = \frac{1}{4}\left(\frac{1}{\xi_1^2} + \frac{1}{\kappa_1^2}\right), \qquad b = \frac{1}{4}\left(\frac{1}{\xi_1^2} - \frac{1}{\kappa_1^2}\right)$$

and $I_0(x)$ is the modified Bessel function of the first kind, and which has the following integral representation:

$$I_0(x) = \frac{1}{\pi} \int_0^{\pi} e^{x\cos\phi} d\varphi.$$

## 4. Dynamics and order density distribution

Evolution equation for the probability amplitude (wavefunction) can be obtained from the following consideration. Assuming that price exists at all times, we have:

$$\int_0^\infty |\psi(s,t)|^2 \, ds = 1.$$

Then if $\psi$ is a differentiable function of time

$$\psi^* \frac{\partial \psi}{\partial t} + \psi \frac{\partial \psi^*}{\partial t} = 0.$$

The most general linear equation satisfying this condition is:

$$i\rho \frac{\partial \psi}{\partial t} = \hat{Q}\psi \tag{10}$$

which means that time-dependent Schrodinger equation in its general form still holds in this problem. Constant $\rho$ has dimensions of [time·$] and determines the degree of phase jitter experienced by the wavefunction at each step in time. We will come back to this important parameter later, but for now we will set $\rho = 1$ for simplicity. Operator $\hat{Q}$ is identified with $\hat{S}$ through trivial solution $\psi = e^{-i\frac{st}{\rho}}$ when $s_{11} = s_{22} = s$ and $s_{12} = 0$. Bringing these considerations together, we have the dynamic equations in their general form:

$$i\frac{d\psi_{ask}}{dt} = s_{11}\psi_{ask} + s_{12}\psi_{bid} \tag{11a}$$

$$i\frac{d\psi_{bid}}{dt} = s_{12}^*\psi_{ask} + s_{22}\psi_{bid} \tag{11b}$$

For constant coefficients this system of equations has the following solution:

$$\psi_{ask}(t) = e^{-i\frac{s_{11}+s_{22}}{2}t}\left\{\left[\cos(\Delta't) - i\frac{s_{11}-s_{22}}{2\Delta'}\sin(\Delta't)\right]\psi_{ask}(0) - i\frac{s_{12}}{\Delta'}\sin(\Delta't)\psi_{bid}(0)\right\} \quad (12a)$$

$$\psi_{bid}(t) = e^{-i\frac{s_{11}+s_{22}}{2}t}\left\{-i\frac{s_{12}^*}{\Delta'}\sin(\Delta't)\psi_{ask}(0) + \left[\cos(\Delta't) + i\frac{s_{11}-s_{22}}{\Delta'}\sin(\Delta't)\right]\psi_{bid}(0)\right\}, \quad (12b)$$

where $\Delta' = \dfrac{\Delta}{2}$.

Since coefficients $s_{ij}$ are stochastic, the solution has to be modelled numerically. Substituting Eqs. (6a-6c) into Eqs. (12a, 12b) we get:

$$\psi_{ask}(t) = e^{-is_{mid}t}\left\{\left[\cos(\Delta't) - i\frac{\xi}{2\Delta'}\sin(\Delta't)\right]\psi_{ask}(0) - i\frac{\kappa}{2\Delta'}\sin(\Delta't)\psi_{bid}(0)\right\} \quad (13a)$$

$$\psi_{bid}(t) = e^{-is_{mid}t}\left\{-i\frac{\kappa}{2\Delta'}\sin(\Delta't)\psi_{ask}(0) + \left[\cos(\Delta't) + i\frac{\xi}{2\Delta'}\sin(\Delta't)\right]\psi_{bid}(0)\right\} \quad (13b)$$

These equations have to be applied to initial conditions step after step, while updating the coefficients $\xi$ and $\kappa$ at each step. It is important to propagate Eqs. (13a, 13b) for a sufficient number of steps to allow them lose the memory of initial conditions.

Since probability amplitude $\psi$ itself cannot be observed, we need to focus on order density, for example $p = |\psi_{ask}|^2$. The immediate value of this quantity also cannot be observed, but it's probability distribution $Q(p)$ can. That probability is determined by the time spent by the asset in "ask" state. For constant coefficients $s_{ij}$ it can be written in the following way:

$$Q(p) \sim \frac{dt}{dp} = \frac{1}{\Delta}\frac{1}{\sqrt{(p_{\max} - p)(p - p_{\min})}},$$
$$(14)$$

where $p_{\min}$ and $p_{\max}$ are the minimum and maximum values of $|\psi_{ask}|^2$. If $\kappa \neq 0$, $\xi = 0$, and at the maximum amplitude Eq. (14) reduces to

$$Q(p) \sim \frac{1}{\sqrt{p(1-p)}}.$$ (15)

This means that $p$ obeys Beta distribution: $p \sim \mathrm{B}\left(\frac{1}{2}, \frac{1}{2}\right)$, so that probability of finding the price in a clean "bid" or "ask" state is substantially higher than finding it in a mixed state. Even though the condition $\xi = 0$ is unlikely to uphold strictly, this effect should still be observed in assets with $\xi << |\kappa|$. For $\kappa \sim \xi$ the effect is washed away and probability distribution should be flat.

Quantities $|\psi_{bid}|^2$ and $|\psi_{ask}|^2$ represent order densities and can be measured from the order book. We can say that if $N_{bid}$ is the current total size of best bid orders and $N_{ask}$ is the current total size of best ask orders, then $\frac{|\psi_{bid}|^2}{|\psi_{ask}|^2} \approx \frac{N_{bid}}{N_{ask}}$, or equivalently

$$|\psi_{bid}|^2 \approx \frac{N_{bid}}{N_{ask} + N_{bid}} \quad \text{and} \quad |\psi_{ask}|^2 \approx \frac{N_{ask}}{N_{ask} + N_{bid}}$$ (16)

These relations provide an interesting insight into probability distribution of the order sizes themselves. As noted above, in case of a large transfer coefficient $\kappa$ probability distribution of $|\psi_{ask}|^2$ follows Beta distribution. From this fact and Eqs. (16) we can deduce that order sizes must obey Gamma distribution: $N_{\left\{\begin{smallmatrix} bid \\ ask \end{smallmatrix}\right\}} \sim \Gamma\left(\frac{1}{2}, \theta\right)$, where $\theta$ can take any value, but has to be the same for both $N_{ask}$ and $N_{bid}$.

This is the time where we have to discuss parameter $\rho$ in more detail. This parameter is responsible for the degree of phase jitter experienced by the wavefunction at each step in time. If $\rho$ is large then phase shift acquired by the wavefunction at each step is much smaller compared to $2\pi$. As a result evolution takes place adiabatically. With small $\rho$ phase shift can be large enough to wash away any possible patterns. Thus, depending on the value of $\rho$ probability distribution $Q(p)$ can change its form. This means that $\rho$ can be found from calibration to order density distribution in the order book data.

## 5. Calibration to market data

*Observables*

Equations Eq. (7a,7b) and Eq. (13a, 13b) provide general form of behavior to the described two-level system. They do not depend on specific selection of the two levels, time or tick intervals, and other parameters. These can be specified depending on the modeled data. For example, one can choose to model the behavior of best bid (BB) and best ask (BA) in an order book, such as in Fig. 1.

| Bid | | Ask | |
|---|---|---|---|
| **Price** | **Size** | **Price** | **Size** |
| 27.83 | 100 | 27.87 | 100 |
| 27.82 | 100 | 27.9 | 100 |
| 27.8 | 200 | 27.95 | 100 |
| 27.79 | 200 | 28.15 | 100 |
| 27.78 | 100 | 28.2 | 100 |

Fig. 1. Sample order book

Then the BB and BA represent the two levels. The difference between them is the actual spread and $N_{\{bid\} \atop \{ask\}}$ are the order sizes at the BB and BA levels.

Alternatively, one could choose to model the effective bid (EB) and effective ask as (EA) for the top $N$ levels of order book, defined as:

$$EB = \frac{\sum_{i=1}^{N} bid_i \cdot bid\,size_i}{\sum_{i=1}^{N} bid\,size_i} \qquad EA = \frac{\sum_{i=1}^{N} ask_i \cdot ask\,size_i}{\sum_{i=1}^{N} ask\,size_i}$$

Sample order book with effective prices and cumulative order sizes is shown in Fig. 2.

| Bid | | | | Ask | | | |
|---|---|---|---|---|---|---|---|
| **Price** | **Size** | **Cum. Size** | **Eff. Price** | **Price** | **Size** | **Cum. size** | **Eff. Price** |
| 27.83 | 100 | 100 | 27.83 | 27.87 | 100 | 100 | 27.87 |
| 27.82 | 100 | 200 | 27.83 | 27.9 | 100 | 200 | 27.89 |
| 27.8 | 200 | 400 | 27.81 | 27.95 | 100 | 300 | 27.91 |
| 27.79 | 200 | 600 | 27.81 | 28.15 | 100 | 400 | 27.97 |
| 27.78 | 100 | 700 | 27.80 | 28.2 | 100 | 500 | 28.01 |

Fig. 2. Sample order book with cumulative order sizes and effective prices.

It is also possible to model the regular high and low levels of the OHLC ticks (Table 1). The spread would correspond to the OHLC tick height (High–Low). However, population probability distribution would cease to make sense in this case.

Table 1. Sample OHLC data

| Date | Open | High | Low | Close |
|---|---|---|---|---|
| 15-Nov-13 | 37.95 | 38.02 | 37.72 | 37.84 |
| 14-Nov-13 | 37.87 | 38.13 | 37.72 | 38.02 |
| 13-Nov-13 | 36.98 | 38.16 | 36.9 | 38.16 |
| 12-Nov-13 | 37.38 | 37.6 | 37.2 | 37.36 |
| 11-Nov-13 | 37.69 | 37.78 | 37.36 | 37.59 |

Let us now apply the theory to specific real-life examples.

***Best bid-ask calibration***

Probability distribution of spread along with the calibrated curve for best bid-ask levels over particular days are shown in Fig. 3 below. Parameters of calibration are given in Table 2. One can notice that INTC and MSFT are characterized by a large $\kappa_0$ and small $\kappa_1$. Due to large volume and relatively low price these tickers demonstrate very small spread that can vary only within a few cents.

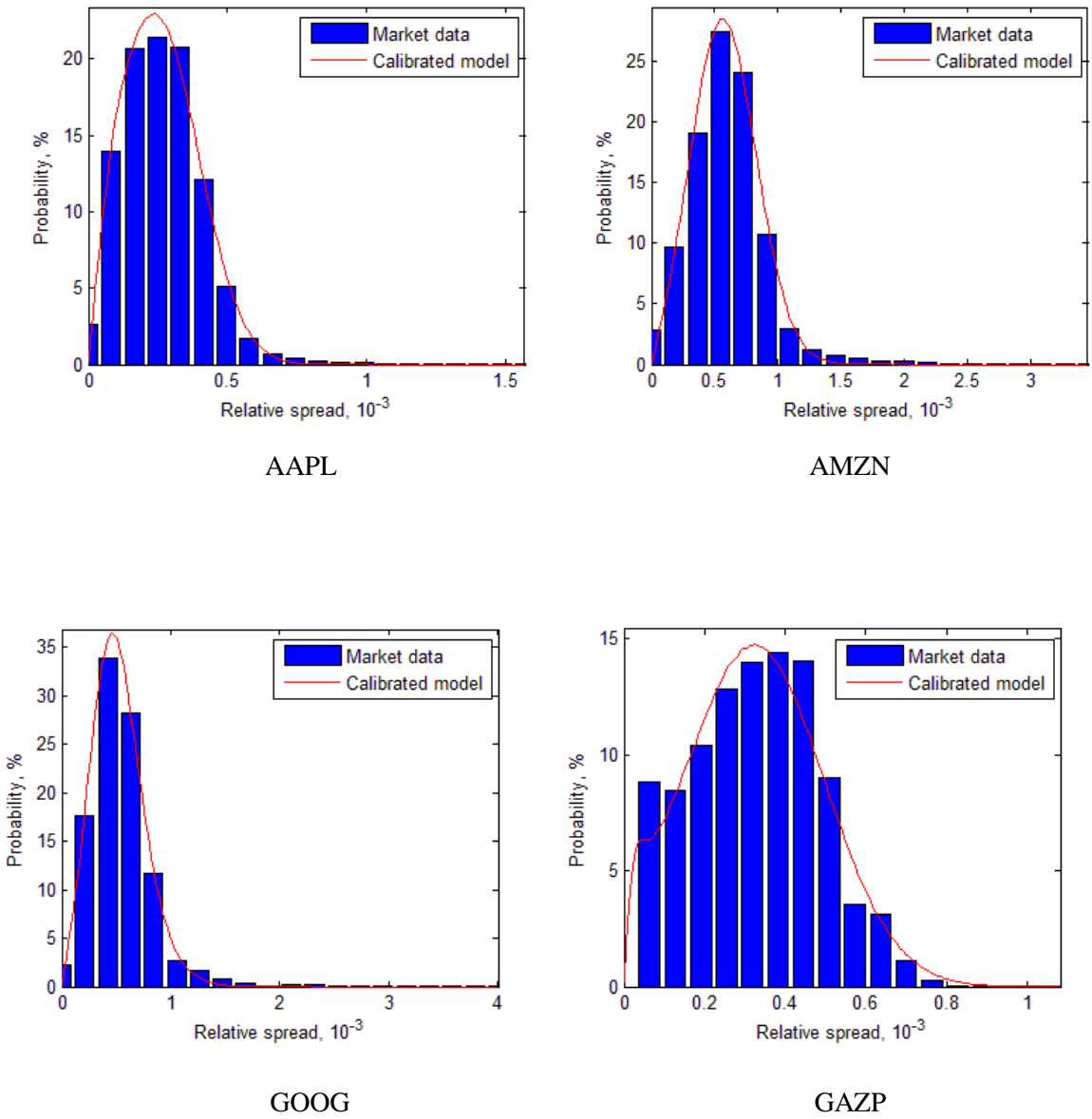

Fig. 3. Probability distribution of spread along with the calibrated curve for best bid-ask levels over particular days

Table 2. Calibration parameters for best bid/ask levels over particular days.

| Ticker | Relative spread - BB/BA $\cdot 10^{-3}$ | | |
|---|---|---|---|
| | $\xi_1$ | $\kappa_0$ | $\kappa_1$ |
| AAPL | 0.11 | 0.23 | 0.16 |
| AMZN | 0.32 | 0.48 | 0.27 |
| GOOG | 0.41 | 0.33 | 0.21 |
| INTC | 0.42 | 0.55 | 0.07 |
| MSFT | 0.26 | 0.32 | 0.016 |
| GAZP (MoEx) | 0.02 | 0.35 | 0.17 |

Another outstanding example is GAZP, traded on Moscow Exchange (MoEx). This "specimen" has low $\xi$ and is therefore supposed to demonstrate the effect of order density inhomogeneity mentioned in Part 4. Charts of order density distribution using parameters from Table 2 plotted against the market data are shown in Fig. 4. One can see that indeed, tails of order density distribution for this stock are curved upward and are about twice larger at the tails than in the middle. It is important to note that charts of Fig. 4 have been generated using the parameters obtained **independently** from the spread data.

One can also observe the characteristic peaks at $p = 0.5$. We tend to think that these peaks have an artificial nature. Orders in highly liquid stocks are mostly submitted in multiples of 100 rather than fractional numbers. As a result, a situation in which 100 bid size is placed against a 100 ask size occurs more frequently than where the ratio is 100 / 70. This causes an inflated frequency of occurrences at $p = 0.5$.

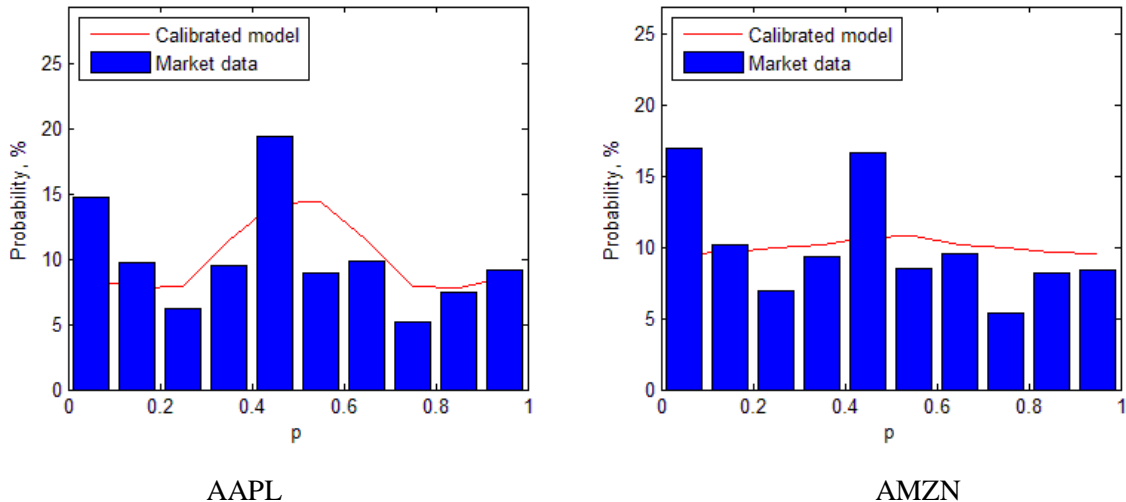

AAPL                                           AMZN

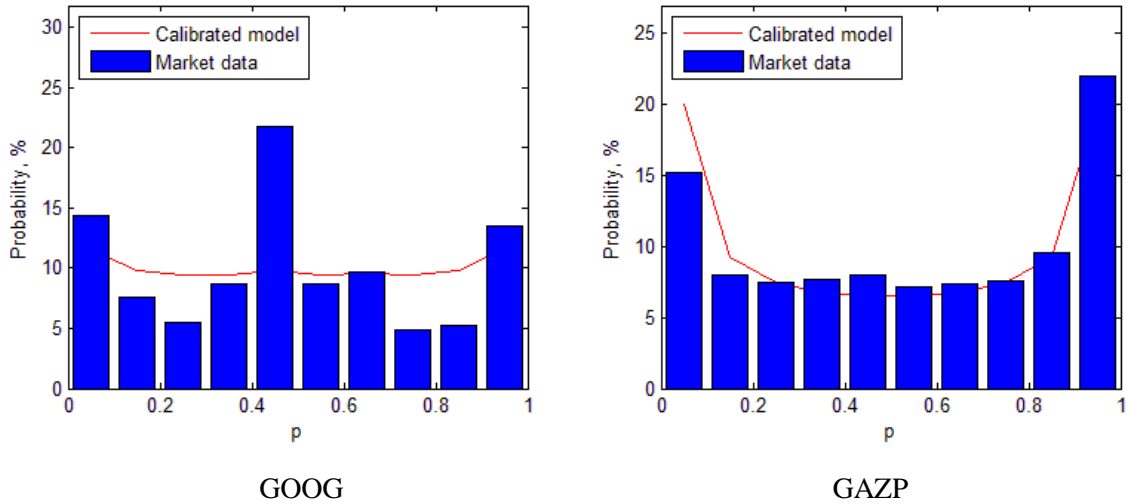

GOOG                          GAZP

Fig. 4. Probability distribution of order density for BB and BA with $\dfrac{s}{\rho}\delta t = 20$. One can see that GAZP

ticker with very low intrinsic component ξ indeed obeys a distribution close to beta B (0.5, 0.5), while all
others don't demonstrate that effect.

### *Tick data calibration*

Similar calculations can be performed with OHLC tick data. In such setup the "high" price would
correspond to "ask" level, and the "low" price would correspond to "bid" level. Calibration results for
relative hi-low are presented in Fig. 5, and calibration parameters are given in Table 3. The order density
notion in this setup should be disregarded, since it has no physical meaning.

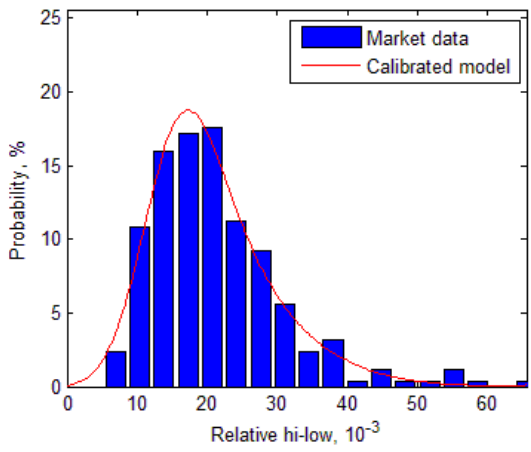

AAPL

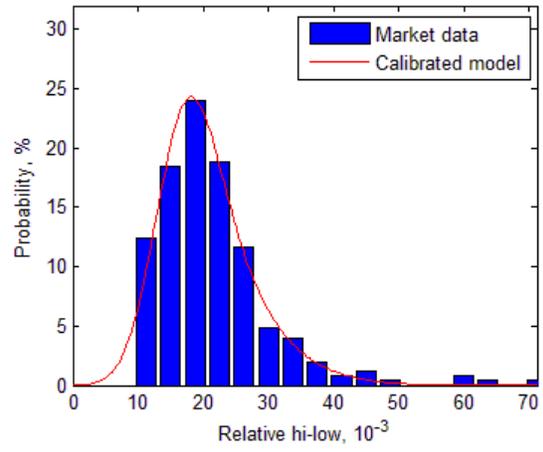

AMZN

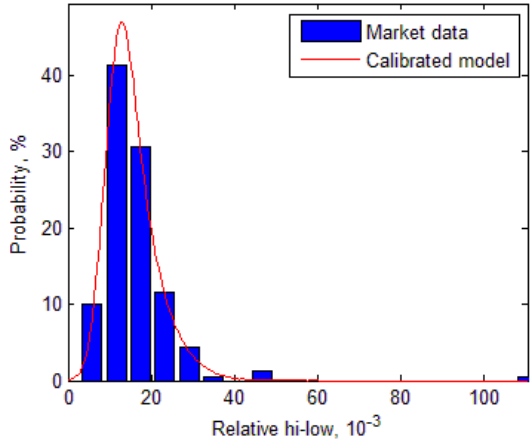

GOOG

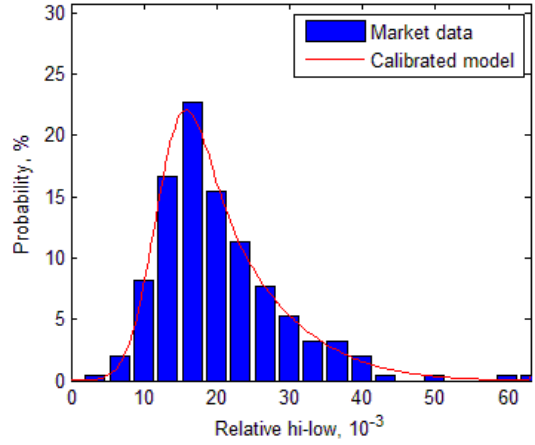

GAZP

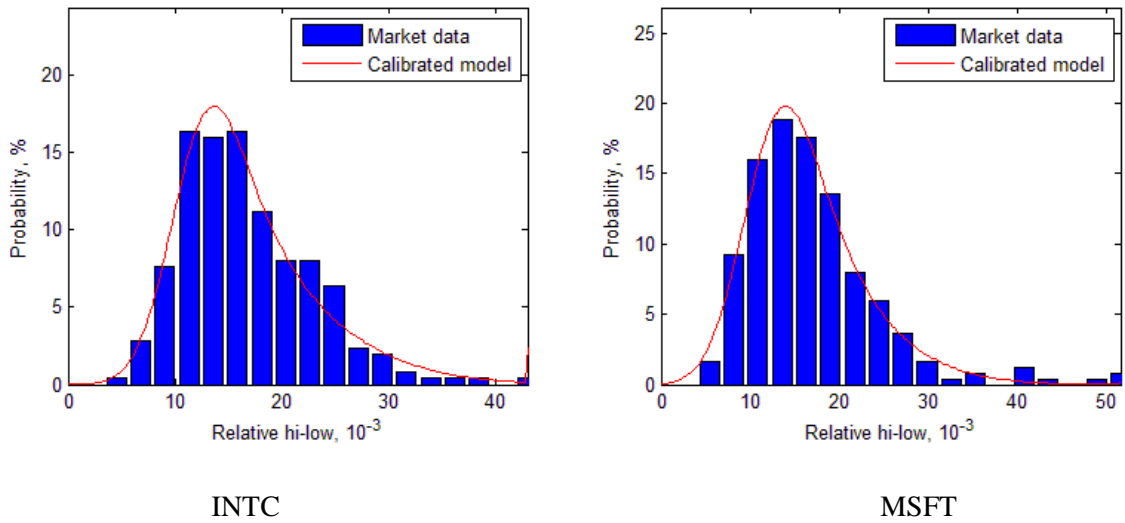

| INTC | MSFT |

Fig. 5. Probability distribution of daily relative hi-low difference for the OHLC ticks against the calibrated curve.

Table 3. Calibration parameters for daily relative hi-low difference for the OHLC ticks.

| Ticker | HI-LOW | | |
|--------|--------|--------|--------|
| | $\xi_1$ | $\kappa_0$ | $\kappa_1$ |
| AAPL | 1.75% | 1.31% | 0.51% |
| AMZN | 1.53% | 1.49% | 0.45% |
| GOOG | 1.24% | 1.02% | 0.34% |
| INTC | 1.39% | 1.11% | 0.31% |
| MSFT | 1.31% | 1.10% | 0.39% |
| GAZP (MoEx) | 1.73% | 1.30% | 0.35% |

## 6. Risk management of assets with limited liquidity

The presented framework provides new capabilities for risk management of assets with limited liquidity. We see that now in addition to traditional "closing price fluctuations" component we can quantify risk arising due to spread and its fluctuations. Overall, risk components can be systematized in Table 4 below. These formulas allow to quantify risk and understand its sources when spread of an asset plays important role in trading activity, particularly in market making.

Table 4. Risk components and corresponding formulas.

| Risk | Meaning | Formula | 95% quantile |
|---|---|---|---|
| Mid-price | Risk of mid-price change | $\sigma = \sqrt{\dfrac{1}{N}\sum_{i=1}^{N}\left(s_{mid,i} - \bar{s}_{mid}\right)^2}$ | $1.65\,\sigma$ |
| Spread | Risk of spread increase | $\Delta = \sqrt{\left(\xi\,du\right)^2 + \left(\kappa\,dv\right)^2}$ | No general analytical expression |

## 7. Discussion

In today's literature one can often find attempts to apply known solutions of quantum mechanical problems to financial markets. True, as a formalism dealing with probabilities it is too tempting not to try such application. Attempts to quantize price, volume, draw an analogue with Heisenberg's uncertainty principle, calculation of transition probabilities, etc, many knows problems of quantum mechanics try to find their way into finance. However, a theory should be based on a concept and allow to draw precise conclusions that are quantifiable, measurable and can be tested in experiment. Simply writing down quantum-mechanical equations and substituting variables with names from finance cannot be the answer. The presented theory is one step in that direction.

Does this mean that price formation has a quantum-mechanical nature? No. We tend to think that quantum mechanical formalism has wider applications than just quantum mechanics and one such area of application is in finance. This is not unusual, since many phenomena in Physics are described by similar equations [10, 11], and financial markets are a physical system too.

Have we missed anything in our analysis? Yes. Our results on order level population density take into account only visible orders. Icebergs, hidden orders, as well as the activity in OTC markets and dark pools was left beyond our description. While these are important factors, we must say a few words why they are also unessential. The described model claims to be conceptual. Therefore, behavior or other layers of trading activity must obey similar laws. Due to the linearity of the model, no layer affects other layers.

The stochastic form of matrix elements in Eqs. (6a-6c) is chosen to facilitate convenient description of market data. We do not claim that this is the only possible choice. Other formats may exist, which may better describe the data. However, we find no evidence that it may be so.

Eigenvalue problem Eq. (1) has been formulated for the price operator. Theoretically, this allows negativity of prices. A more appropriate formulation would have been for the logarithmic price, which would exclude possibility of negative prices and include scaling. In this paper we deliberately based our formulation on price to make ideas easier to grasp.

Strictly speaking, use of wavefunctions in the model suggests that there may be interference between different eigenstates. The fact of its direct occurrence is unknown. However, order book data with GAZP (Eq.(15) and Fig. (4)) indicates that oscillatory behavior may exist in processes associated with that stock. Additional research is required before we can answer that question.

In our model we used volatility $\sigma$ as a separate factor. Clearly, spread $\Delta$ is connected to volatility. This prompts a question: is $\sigma$ required as a separate factor, or can the entire model be formulated using only parameters $\xi$ and $\kappa$.

## 8. Conclusions

Summarizing, we have developed a conceptual framework that provides new capabilities for financial institutions that are involved in market-making and securities dealing activities. This framework allows to model bid and ask prices in a consistent way and we have shown that behaviors resulting from this framework agree to the extent possible in finance with measurable market data.

This model can be calibrated to various types of data, such as best bid-and-ask, effective bid-ask, or even OHLC ticks. Using the calibrated model one can measure risk associated with spread, gauge it against the regular mid-price risk, calculate the possible range for bid and ask prices at the end of time horizon, etc. All these capabilities are extremely important when a trading desk's risk/return profile depends substantially on spread.

This model also opens opportunities for new research, such as dynamics of eigenstates, option pricing and interaction with external factors.